\documentclass[
  preprint,
  superscriptaddress,
  amsmath,amssymb,
  aps,
  prapplied,
  endfloats 
]{revtex4-2}

\usepackage{newtxtext,newtxmath} 
\usepackage{graphicx}            
\usepackage{dcolumn}             
\usepackage{bm}                  

\usepackage[mathlines]{lineno}

\PassOptionsToPackage{hidelinks}{hyperref}
\usepackage{hyperref}

\begin{document}

\title{Effect of Magnetic Anisotropy on Magnetoelastic Waves in Ni/LiNbO$_3$ Hybrid Device}

\author{Minwoo Yu}
\thanks{These authors contributed equally to this work.}
\affiliation{Department of Physics, Pohang University of Science and Technology, Pohang 37673, Korea}

\author{Moojune Song}
\thanks{These authors contributed equally to this work.}
\affiliation{Department of Physics, Korea Advanced Institute of Science and Technology (KAIST), Daejeon 34141, Korea}

\author{Minseok Kang}
\affiliation{Department of Materials Science and Engineering, KAIST, Daejeon 34141, Korea}

\author{Mujin You}
\affiliation{Department of Physics, Korea Advanced Institute of Science and Technology (KAIST), Daejeon 34141, Korea}

\author{Yunyoung Hwang}
\affiliation{Institute of Innovation for Future Army, Daejeon, Republic of Korea Army}

\author{Albert Min Gyu Park}
\affiliation{Department of Physics, Korea Advanced Institute of Science and Technology (KAIST), Daejeon 34141, Korea}

\author{Byong-Guk Park}
\affiliation{Department of Materials Science and Engineering, KAIST, Daejeon 34141, Korea}

\author{Kab-Jin Kim}
\affiliation{Department of Physics, Korea Advanced Institute of Science and Technology (KAIST), Daejeon 34141, Korea}

\author{Junho Suh}
\thanks{Contact author: junhosuh@postech.ac.kr}
\affiliation{Department of Physics, Pohang University of Science and Technology, Pohang 37673, Korea}

\begin{abstract}
We study the effects of magnetic anisotropy and crystalline axes in surface acoustic waves (SAWs) driven magnetic resonances of Ni/$\text{LiNbO}_3$ hybrid devices.~SAW absorption from the interaction with magnons in Ni displays a strong anisotropic dependence on the direction of the applied in-plane magnetic field.~Magnetic anisotropy is further investigated by magneto-optical Kerr effect measurements to show both uniaxial and biaxial anisotropy components in Ni films on $\text{LiNbO}_3$.~By introducing a dipolar interaction term in addition to the anisotropies, we successfully explain the anisotropic SAW absorption in our devices.~These findings show the importance of substrate-induced anisotropy and long-range dipolar effects in SAW–magnon hybrid devices and indicate future directions for optimizing these spin-acoustic devices through comprehensive anisotropy engineering.
\end{abstract}

\maketitle

\section{\label{sec:level1}INTRODUCTION}

Hybrid quantum systems, based on strong coupling between disparate excitations, can achieve advantages or functionalities that are unattainable with a single excitation~\cite{Kurizki2015,Gustafsson2014,Chu2017,Satzinger2018,LachanceQuirion2019,Clerk2020}.~In particular, magnon–phonon systems, which couple magnons—the quanta of collective spin excitations (spin waves)—and phonons—the quanta of lattice vibrations (acoustic waves)—enable novel functionalities such as the improvement of magnon propagation leveraging the low damping property of phonons~\cite{Hioki2022}, and the generation of phonon nonreciprocity induced by angular momentum in magnetic materials~\cite{Xu2020,Sasaki2021,Liao2024}. 

Hybrid systems with magnetic thin films, such as Ni or CoFeB on LiNbO$_3$, have been widely investigated~\cite{Xu2020,Sasaki2021,Weiler2011,Hwang2024,Kuss2021,Kuss2023,Matsumoto2022a,Matsumoto2024b}.~These structures offer relatively easy fabrication and demonstrate strong magnon–phonon coupling in microwave-frequency measurements.~In these examples, surface acoustic wave (SAW) phonons propagating through LiNbO$_3$ hybridize with spin waves in the magnetic film. Nickel has been utilized in studies of magnon–phonon hybrid systems due to its large magnetoelastic coefficient, placing it as a promising platform since the early demonstrations of magnon–phonon coupling with SAWs~\cite{Weiler2011}.~In addition, magnetized Ni films exhibit strong optical responses in magneto-optical Kerr effect (MOKE) and Brillouin light scattering (BLS) microscopies, making these techniques ideal for visualizing magnetoelastic waves induced by magnon–phonon interactions~\cite{Zhao2021,Rovirola2024,Komiyama2024,Kimel2022,Kunz2024}. 

Previous experiments on Ni/LiNbO$_3$ devices have demonstrated that the magnetoelastic coupling between Rayleigh-mode SAWs and spin waves is proportional to $\sin{2\varphi_h}$, where $\varphi_h$ is the angle between the SAW propagation direction and the applied in-plane magnetic field, and it is maximized when $\varphi_h = 45^{\circ}$~\cite{Weiler2011}.~However, several studies have reported deviations from this, with maximum coupling around $\varphi_h = 30^{\circ}$, attributing this discrepancy to magnetic anisotropy within the Ni films on LiNbO$_3$~\cite{Hatanaka2022,Gao2022}.~These earlier studies assumed a simple uniaxial anisotropy without thorough consideration of its origin or relationship to the Ni/LiNbO$_3$ interface. 

In this study, we experimentally demonstrate that the characteristics of magnon–phonon coupling in Ni/LiNbO$_3$ hybrid devices depend on the SAW propagation direction relative to the LiNbO$_3$ crystal axes, establishing a clear connection between coupling phenomena and substrate crystal orientation.~Measurements using MOKE show that Ni films deposited on LiNbO$_3$ exhibit biaxial magnetic anisotropy, which is closely related to the crystal orientation of the LiNbO$_3$ substrate. These observations suggest that the crystal orientation of the LiNbO$_3$ substrate can induce anisotropic strain at the interface, which in turn shapes the magnetic anisotropy of the Ni film~\cite{Ito2022}.~The uniaxial and biaxial magnetic anisotropies are identified as essential to explain the magnetic-field dependence of SAW absorption of our Ni/LiNbO$_3$ device in addition to a dipolar interaction. Our findings highlight how interface-driven magnetic properties influence magnon–phonon interactions in Ni/LiNbO$_3$ systems, emphasizing crucial considerations for future research into on-chip magnon–phonon hybrid devices.

\section{\label{sec:level2}EXPERIMENTAL SETUP}

We fabricate two-port SAW resonators on a 128$^{\circ}$ Y-cut LiNbO$_3$ substrate with two different SAW propagation directions—one along the crystallographic X axis (Device~1) and the other along the Y$'$ axis, perpendicular to the X axis (Device~2).~The transmission from port~2 to port~1 ($S_{12}$) was measured using a vector network analyzer (VNA) for each device [Fig.~1(a)].

\begin{figure}[t]
    \centering
    \includegraphics[width=1\linewidth]{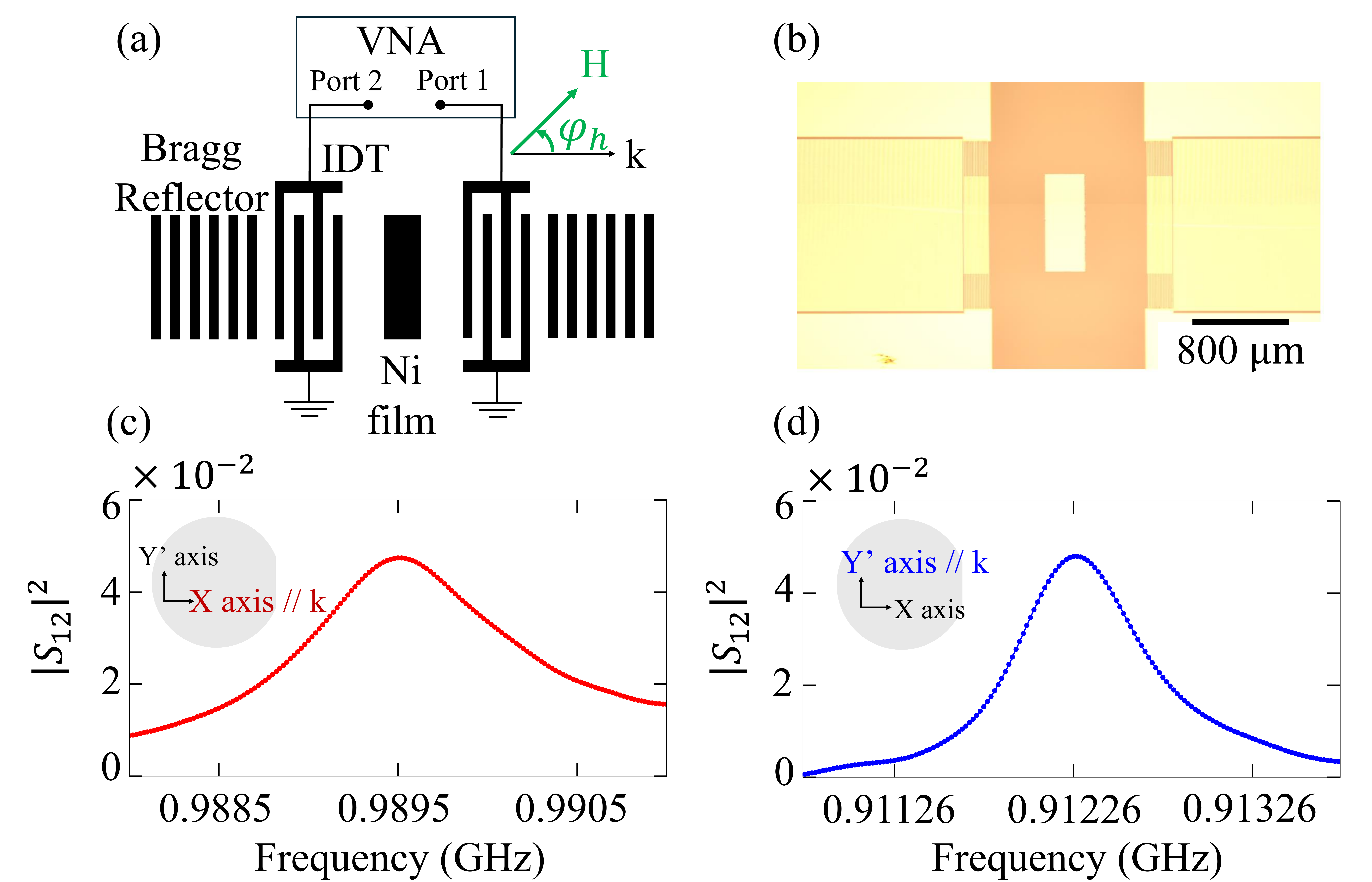}
    \caption{(a) Schematic top view of the device structure and
    experimental setup for magnetostrictive coupling in a Ni film
    and surface acoustic waves (SAWs) on a black LiNbO$_3$ substrate.
    (b) Optical image of Device~1.~(c),(d) Frequency dependence of
    SAW transmission, $|S_{12}|^2$, along X-axis propagation and
    Y$'$-axis propagation, respectively.}
    \label{fig:device}
\end{figure}
The interdigital transducer (IDT) and the Bragg reflector are made up of 50 nm thick aluminum stripes with a width and spacing of $1\,\mu\mathrm{m}$, respectively. At the center of the two IDTs, a 20-nm thick Ni film was deposited by dc magnetron sputtering with a 3-nm thick Ta capping layer to prevent Ni oxidation [Fig.~1(b)].~Figures~1(c) and 1(d) show the VNA spectra of Device~1 and Device~2, respectively, with enlarged views around the SAW resonance peaks. 

The difference in resonance frequencies between the two devices originates from the distinct phase velocities of the SAWs propagating along the X and Y' axes of the LiNbO$_3$ substrate~\cite{Liao2024,Hashimoto2000,Weser2020,Yamada1987}.~Here, each SAW resonance peak was selected from among multiple resonance peaks generated by the Bragg mirror resonator, specifically choosing the peak with the maximum SAW transmission. The measured linewidth for Device~1 and Device~2 are $1.38\,\mathrm{MHz}$ and $0.84\,\mathrm{MHz}$, corresponding to quality factors of $Q = 715$ and $Q = 1080$, respectively.

\section{\label{sec:level3}MAGNON–PHONON COUPLING MEASUREMENT}
We measure the effects of magnetoelastic coupling between SAWs and magnons in Ni by recording $S_{12}$ at one of the SAW resonant frequencies under external magnetic fields oriented from $0^{\circ}$ to $180^{\circ}$ relative to the SAW propagation direction ($k_{\mathrm{SAW}}$). The SAW frequencies of Device~1 and Device~2 are $0.9895\,\mathrm{GHz}$ and $0.91246\,\mathrm{GHz}$, respectively. Figure~2 shows the absorption of SAWs in a Ni thin film under an in-plane configuration, illustrating variations as a function of the magnetic-field strength and its orientation~\cite{Weiler2011,Hwang2024,Hatanaka2022,Gao2022}.  

We define the normalized SAW transmission as
\[
\Delta\left|S_{12}\right|^2 = \frac{\left|S_{12}\right|^2}{\left|S_{12}\right|_{\mathrm{ref}}^2},
\]
where $\left|S_{12}\right|_{\mathrm{ref}}^2$ is the value at $\mu_{0}H = 10\,\mathrm{mT}$. To examine these interactions more closely, we extract slices of the SAW transmission data at specific magnetic-field angles ($15^{\circ}$, $45^{\circ}$, and $150^{\circ}$), as shown in Figs.~2(c)–2(h). 

Device~1 shows double dips in the SAW transmission when the external magnetic field is applied near $45^{\circ}$ relative to the direction of $k_{\mathrm{SAW}}$, whereas only a single dip appears at $0^{\circ}$ and $90^{\circ}$, as shown in the Appedix A.~This behavior indicates SAW-driven spin-wave resonance consistent with previous studies~\cite{Hatanaka2022}.~According to Ref.~[22], the magnetoelastic coupling is maximized when the magnetization angle.~$\varphi_{m}$ satisfies
\begin{equation}
\varphi_{m} = 45^{\circ} + 90^{\circ} \times n \quad (n = 0, 1, 2, \dots).
\end{equation}
When the external magnetic field is applied at $45^{\circ}$, the initial condition $\varphi_{m} = \varphi_{h}$ leads to the maximum magnetoelastic coupling. As a result, the dip width at $45^{\circ}$ is larger than those at $0^{\circ}$ and $90^{\circ}$. In other words, the magnetoelastic coupling is initially zero when $\varphi_{h} = 0^{\circ}$ or $90^{\circ}$.~As the external magnetic field decreases, the effect of magnetic anisotropy dominates over the Zeeman effect. Consequently, $\varphi_{m}$ shifts toward $45^{\circ}$ or $225^{\circ}$, where the coupling is maximized, resulting in an efficient magnetoelastic interaction. Thus, even for $\varphi_{h} = 0^{\circ}$ and $90^{\circ}$, the magnetoelastic coupling remains nonzero, producing measurable absorption in the SAW transmission. 

In contrast, the dip at $150^{\circ}$ appears broader than at $135^{\circ}$, as shown in Fig.~2(a). Furthermore, Device~2 does not show a double dip near $45^{\circ}$, deviating from the expected trend; instead, the double-dip feature appears near $15^{\circ}$. This behavior suggests a distinct anisotropic response, likely arising from structural variations or crystalline anisotropy.~Our results indicate that an additional anisotropy mechanism modifies the SAW–magnon coupling behavior and highlight the need to consider such effects for a complete description.

\begin{figure*}
\includegraphics[width=\linewidth]{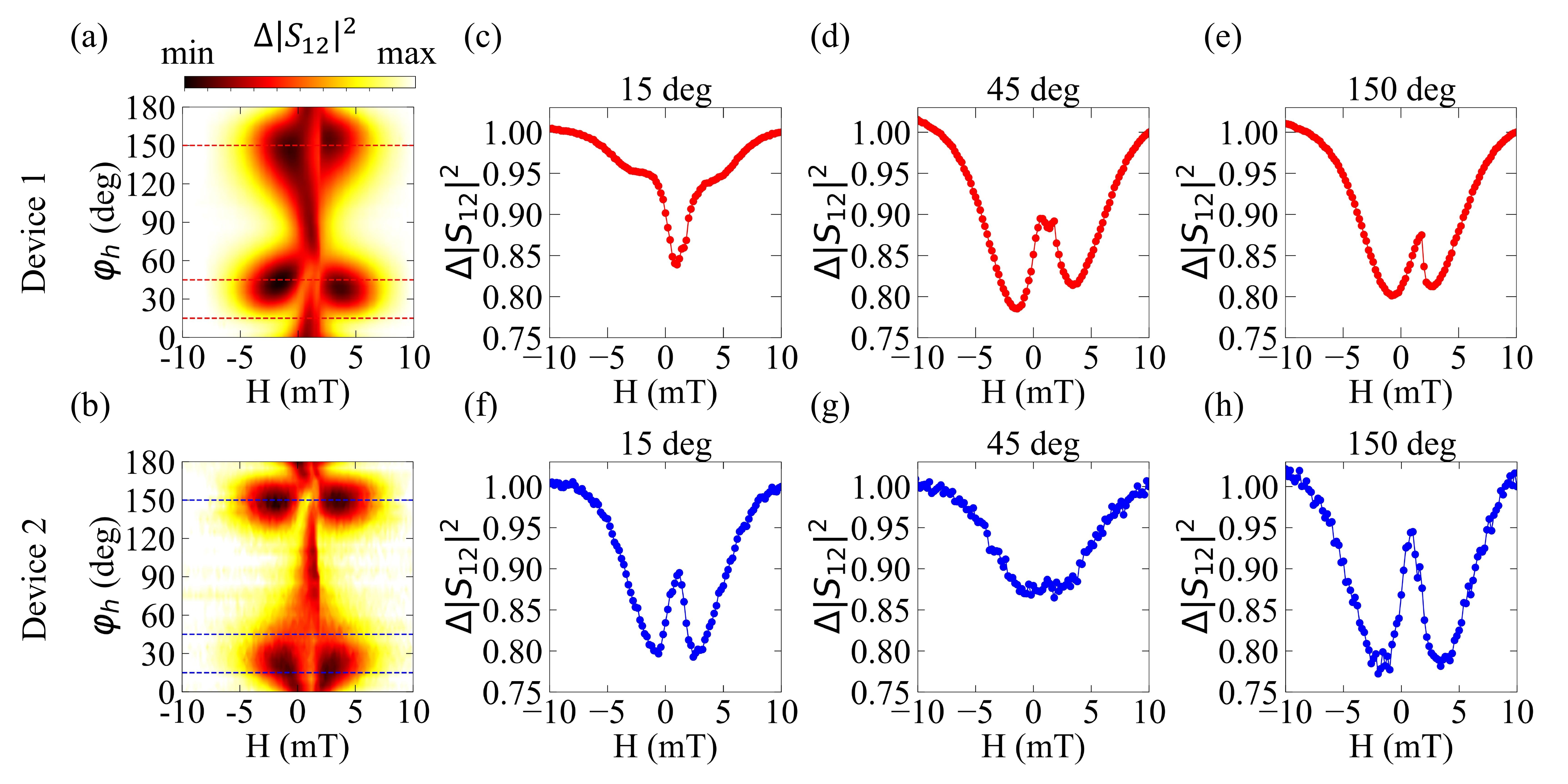}
\caption{(a),(b) Normalized SAW transmission as a function of the magnitude 
        and orientation of the magnetic field, measured at the SAW resonance frequencies 
        of 0.9895~GHz (Device~1) and 0.91246~GHz (Device~2). 
        (c)--(h) Magnetic-field dependence of SAW transmission at 
        $\varphi_{h} = 15^{\circ}$, $45^{\circ}$, and $150^{\circ}$ for Device~1 (c--e) 
        and Device~2 (f--h).}
\end{figure*}

\section{\label{sec:level4}MAGNETO-OPTICAL KERR EFFECT MEASUREMENTS}

We investigate the magnetic anisotropy in our Ni films using the magneto-optical Kerr effect (MOKE). At various angles between the SAW propagation direction and the magnetic-field direction, we sweep the magnetic-field strength to extract the magnetic coercivity from the hysteresis loop.

The polar plots in Figs.~3(a) and 3(b) show the magnetic-field dependence of the coercivity, which exhibits a combination of twofold and fourfold symmetric dependence.~We can see clear existence of the fourfold symmetric component by representing angle-dependence of coercivity $H_{c}(\theta)$ with a fitting curve,
\begin{equation}
H_{c}(\theta) = K_{0} + K_{1} \sin\left( 2\theta + \varphi_{1} \right) + K_{2} \sin\left( 4\theta + \varphi_{2} \right),
\end{equation}
where $K_{0}$, $K_{1}$, and $K_{2}$ correspond to a constant offset, a twofold, and a fourfold symmetric component, respectively, with the associated easy-axis angles $\varphi_{1}$ and $\varphi_{2}$~\cite{Hubert1998}. 

These angle-dependent coercivities from the two devices indicate differences in the separation angle between the easy axis (EA) and hard axis (HA): the EA–HA separation is $120^{\circ}$ for Device~1 and $100^{\circ}$ for Device~2, both differing from the $90^{\circ}$ expected with uniaxial anisotropy. Note that in Figs.~3(a) and 3(b), $0^{\circ}$ is defined with respect to the $k_{\mathrm{SAW}}$ direction for both devices.~This observation confirms the presence of biaxial magnetic anisotropy in both devices.~The anisotropies, possibly induced by interfacial strain due to the crystallographic orientation of the LiNbO$_3$ substrate, modify the magnetic free-energy landscape.~This altered energy landscape shifts the equilibrium position of $\varphi_{m}$; therefore, both uniaxial and biaxial anisotropy should be considered for the accurate description of SAW–magnon interactions in our devices.

\begin{figure*}
\includegraphics[width=\textwidth]{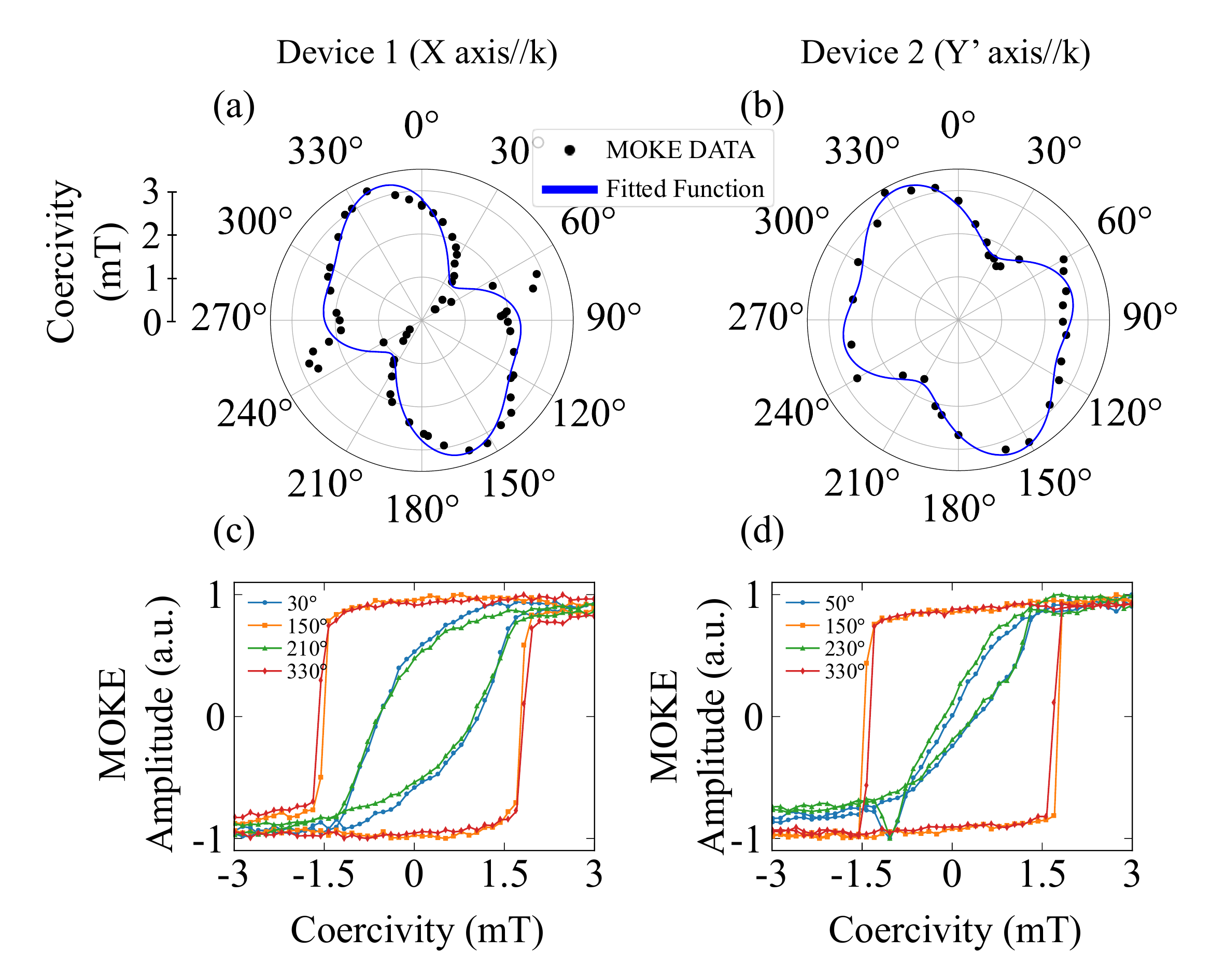}
 \caption{(a),(b) Polar plots of coercivity extracted from hysteresis loops 
measured at various magnetic-field angles for SAW propagation in patterned 
Ni films in Device~1 and Device~2, respectively. 
(c),(d) Hysteresis loops measured at the easy-axis (E.A.) and hard-axis (H.A.) 
angles for each device, respectively.}
\end{figure*}

\section{\label{sec:level5}CALCULATION}

To describe the equilibrium magnetization state of Ni, it is necessary to determine the angle $\varphi_{m}$ that minimizes the free-energy density $G$, which governs the competition between the external magnetic field and the intrinsic anisotropy energy of the ferromagnetic material.~Conventionally, the Stoner–Wohlfarth model describes the magnetization of a single-domain ferromagnet under uniaxial anisotropy~\cite{Hatanaka2022,Gao2022}.~However, our MOKE measurements indicate that Ni films on LiNbO$_3$ exhibit biaxial magnetic anisotropy.~Thus, we extend the Stoner–Wohlfarth model to include biaxial anisotropy terms. We also find that the exchange and dipolar interaction terms are necessary to explain the details of our SAW absorption data. Our model reads
\begin{equation}
\begin{split}
G = & - H \cos\left( \varphi_{h} - \varphi_{m} \right) 
  + B_{u1} \sin^{2} \left( \varphi_{m} - \varphi_{1} \right)  \\
  & + B_{u2} \sin^{4} \left( \varphi_{m} - \varphi_{2} \right) 
  + \frac{A}{M_{s}} k^{2}  \\
  & + \frac{\mu_{0} M_{s}}{2} 
    \left( \frac{1 - e^{-k d}}{k d} \right),
\end{split}
\end{equation}
where $H$ is the magnitude of the external magnetic field, $\varphi_{h}$ is the field direction, $B_{u1}$ and $B_{u2}$ correspond to the strengths of uniaxial and biaxial anisotropy with easy-axis orientations $\varphi_{1}$ and $\varphi_{2}$, respectively, and the fourth and fifth terms represent the exchange and dipolar energy contributions with saturation magnetization $M_{s}$, exchange stiffness $A$, spin-wave wavenumber $k$, and film thickness $d$~\cite{Hwang2024}. 

This biaxial extension introduces additional energy minima, modifying the equilibrium magnetization behavior beyond what is predicted by the conventional Stoner–Wohlfarth model. To validate this theoretical framework, we find the equilibrium position of the magnetization numerically through energy minimization, and implement the Landau–Lifshitz–Gilbert (LLG) equation to model the dynamic response. The LLG equation describes the time evolution of the magnetization $\vec{M}$ driven by an effective magnetic field $\vec{H}_{\mathrm{eff}}$~\cite{Landau1935,Gilbert2004}:
\begin{equation}
\frac{\partial \vec{M}}{\partial t} 
  = - \gamma \, \vec{M} \times \vec{H}_{\mathrm{eff}} 
  + \alpha \, \vec{M} \times \frac{\partial \vec{M}}{\partial t},
\end{equation}
where $\gamma$ is the gyromagnetic ratio and $\alpha$ is the Gilbert damping coefficient. Here, $\vec{H}_{\mathrm{eff}}$ includes contributions from the external field, magnetic anisotropy, and magnetoelastic coupling effects. 

Within the LLG framework, the influence of the strain-induced magnetoelastic field on magnetization dynamics can be analyzed, including its role in energy dissipation and resonance behavior. The absorbed power $P$ provides a quantitative measure of the energy transfer between SAWs and spin dynamics, directly linking the magnetoelastic interaction to spin-wave excitation:
\begin{equation}
P = \frac{\omega \mu_{0}}{2} \, \mathrm{Im} \left( h_{\mathrm{me}}^{T} \cdot \chi \cdot h_{\mathrm{me}} \right) V,
\end{equation}
where $\omega$ is the angular frequency, $\mu_{0}$ is the permeability, and $V$ is the volume of the ferromagnetic thin film~\cite{Weiler2011}.~The magnetoelastic field $h_{\mathrm{me}}$ couples to the magnetic susceptibility tensor $\chi$, which describes the system’s response to external excitations~\cite{Gao2022,Dreher2012,Tateno2020,Gao2023}.~The detailed expressions for the susceptibility components, including coefficients such as $G_{11}$, $G_{22}$, and $G_{3}$, are provided in the Appendix B. The imaginary component of this interaction determines the dissipated energy, providing insight into SAW-driven spin-wave resonance. 

\begin{figure*}
\includegraphics[width=0.7\linewidth]{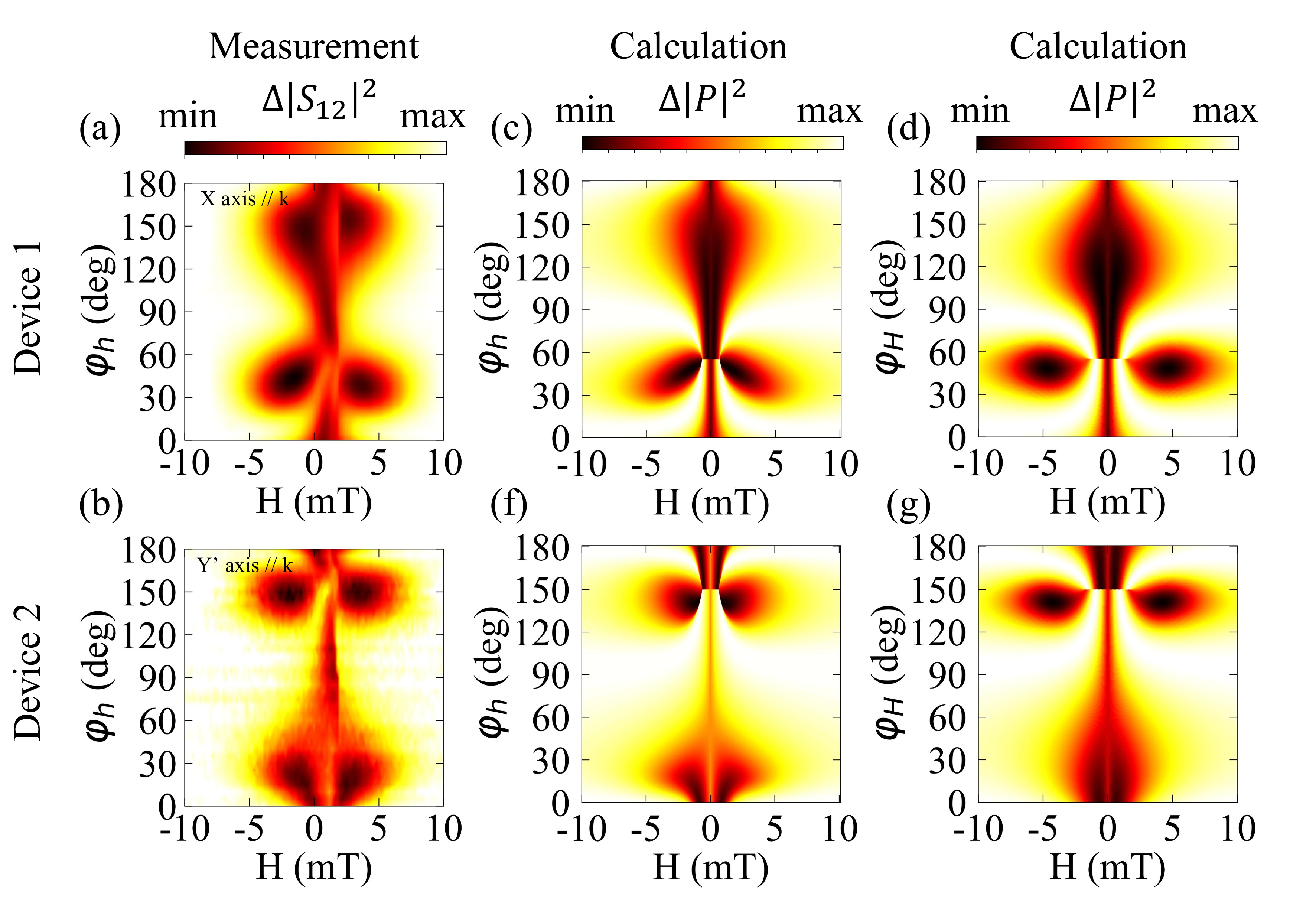}
\caption{(a),(b) Experimental results for Device~1 and Device~2, respectively. 
(c),(f) Calculation results for Device~1 and Device~2, respectively, 
considering biaxial anisotropy and dipolar interaction. 
(d),(g) Calculation results for Device~1 and Device~2, respectively, 
considering only uniaxial anisotropy, where only $B_{u1}$ and 
$\varphi_{\mathrm{EA1}}$ were included in the calculations.}
\end{figure*}
The calculation results based on this model are compared with experimental data to evaluate the influence of biaxial anisotropy and dipolar interaction on magnon–phonon coupling behavior. Figure~4 compares the experimental results for Device~1 and Device~2 with calculations incorporating uniaxial and biaxial anisotropy terms as well as the dipolar interaction term, presenting the normalized absorption as
\[
\Delta\left|S_{12}\right|^{2} = \frac{\left|S_{12}\right|^{2}}{\left|S_{12}\right|_{10\,\mathrm{mT}}^{2}}, 
\quad
\Delta\left|P\right|^{2} = \frac{\left|P\right|^{2}}{\left|P\right|_{10\,\mathrm{mT}}^{2}}.
\]
The parameters used in the calculations are summarized in Appendix C. The selected values of $B_{u1}$, $B_{u2}$, $\varphi_{1}$, and $\varphi_{2}$ provide the best reproducibility of the experimental results and are in good agreement with previous reports~\cite{Hwang2024,Hatanaka2022,Gao2022}.

\section{\label{sec:level6}DISCUSSION}

Our experimental results reveal distinct absorption features that cannot be fully explained by a simple uniaxial anisotropy model. ~The comparison with calculations demonstrates that incorporating both biaxial anisotropy and dipolar interactions is essential for accurately reproducing the observed SAW–magnon coupling behavior. For Device~1, the experimental results in Fig.~4(a) exhibit strong absorption centered around $45^{\circ}$ and $150^{\circ}$, which are well captured in our calculation in Fig.~4(c). In particular, the calculation successfully reproduces the characteristic feature observed near $45^{\circ}$, where the resonant field decreases progressively with increasing angle, reflecting the angular dependence of the absorption shape. This agreement confirms that the energy landscape governing the magnon–phonon interaction is shaped not only by twofold and fourfold anisotropy components but also by dipolar interactions, all of which modify the resonance conditions. 

In contrast, the uniaxial model in Fig.~4(d) fails to reproduce these characteristic absorption features, indicating that an additional anisotropic contribution beyond simple twofold symmetry is essential. Similarly, for Device~2, the absorption behavior in Fig.~4(b) shows dominant features around $30^{\circ}$ and $150^{\circ}$, which are well reproduced in the calculation including biaxial anisotropy and dipolar interactions in Fig.~4(f). The failure of the uniaxial model in Fig.~4(g) to predict these absorption regions further highlights the necessity of incorporating both biaxial anisotropy and long-range dipolar effects into the free-energy expression. The presence of these distinct absorption angles suggests that the SAW propagation direction modifies the effective magnetic-energy landscape, resulting in angle-dependent variations in the magnon–SAW coupling strength depending on the crystallographic orientation of LiNbO$_3$. These results reinforce that both biaxial anisotropy and dipolar interactions are essential to explain the angular dependence of magnon–phonon interactions in Ni/LiNbO$_3$ hybrid devices.

\section{\label{sec:level7}CONCLUSION}

We investigate the interaction between SAWs and spin waves in Ni thin films and identify the role of magnetic anisotropy influenced by the crystallographic orientation of the LiNbO$_3$ substrate.~Our results confirm that SAW transmission characteristics are governed by anisotropy-driven magnetoelastic coupling, which depends on both the magnitude and orientation of the external magnetic field.~MOKE measurements reveal deviations from the expected uniaxial anisotropy, highlighting the significant role of biaxial contributions in shaping SAW–magnon interactions.~Furthermore, the inclusion of both biaxial anisotropy and dipolar interactions in the calculation enables a detailed reproduction of the experimental absorption features, particularly their angular dependence.~These results demonstrate that both substrate-induced biaxial anisotropy and long-range dipolar effects play critical roles in determining magnetization dynamics and must be considered in theoretical descriptions of SAW–magnon coupling.~Our findings provide insights for future studies on SAW–magnon interactions, especially regarding the influence of the SAW propagation direction and anisotropy configuration~\cite{Liao2024,Kawada2024}.~Further investigations into excitation conditions and geometrical effects will be essential for advancing on-chip quantum technologies and magnon-based information processing platforms.

\begin{acknowledgments}
This work was supported by the National Research Foundation of Korea (NRF) and the Institute of Information \& Communications Technology Planning \& Evaluation (IITP), funded by the Ministry of Science and ICT (RS-2023-00207732, RS-2024-00352688, RS-2024-00402302, RS-2023-00259676, RS-2022-00164799).
\end{acknowledgments}

\appendix 
\section{SAW TRANSMISSION DEPENDENCE ON MAGNETIC FIELD AND ANGLE}

\begin{figure}[h] 
    \centering
    \includegraphics[width=0.4\linewidth]{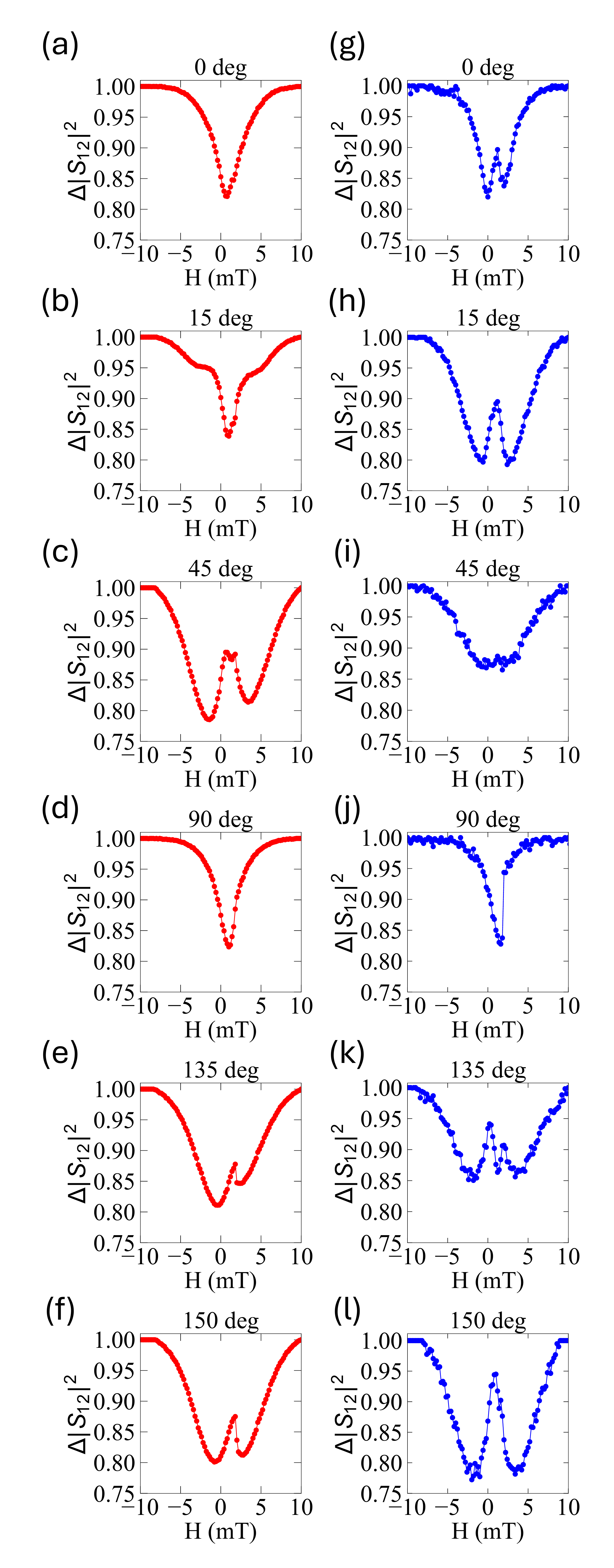}
    \caption{(a-f) Magnetic field dependence of SAW transmission for device 1; (g-l) corresponding measurements for device 2.}
    \label{fig:appendix_transmission}
\end{figure}

Figure~\ref{fig:appendix_transmission} shows the dependence of SAW transmission on the external magnetic field orientation. Fig.~\ref{fig:appendix_transmission}(a–f) present the results for Device~1, and Fig.~\ref{fig:appendix_transmission}(g–l) show the corresponding data for Device~2. The angular dependence of the transmission is different between the two devices.

\section{DERIVATION OF THE EXTENDED FREE-ENERGY MODEL}

From the work of Dreher \textit{et al.}~\cite{Dreher2012}, the normalized free energy is expressed as
\begin{align}
G &= - \mu_{0} \mathbf{H} \cdot \mathbf{m}
   + B_{d} m_{z}^{2}
   - B_{u} (\mathbf{m} \cdot \mathbf{u})^{2},
\label{eq:S1}
\end{align}
where $\mathbf{H}$ is the external magnetic field, $\mathbf{m} = (m_{x}, m_{y}, m_{z})^{T}$ is the unit vector of magnetization, $B_{d}$ is the shape anisotropy, $B_{u}$ is the uniaxial in-plane anisotropy, and $\mathbf{u}$ is the unit vector of the uniaxial anisotropy.  
The exchange energy is neglected here because of the small wavenumber in our magnon–phonon coupling system.

In our case, the Ni film exhibits biaxial anisotropy. Let us consider two in-plane anisotropy fields $B_{u1}$ and $B_{u2}$ with corresponding unit vectors $\mathbf{u}_{1}$ and $\mathbf{u}_{2}$. Then, the normalized free energy becomes
\begin{align}
G &= - \mu_{0} \mathbf{H} \cdot \mathbf{m}
   + B_{d} m_{z}^{2}
   - B_{u1} (\mathbf{m} \cdot \mathbf{u}_{1})^{2} \nonumber\\
  &\quad - B_{u2} (\mathbf{m} \cdot \mathbf{u}_{2})^{4}
   + \frac{A}{M_{s}} k^{2} \nonumber\\
  &\quad + \frac{\mu_{0} M_{s}}{2} 
     \left[ \frac{1 - e^{-k d}}{k d} \cos^{2} \theta \right. \nonumber\\
  &\quad\quad
     \left. + \left( 1 - \frac{1 - e^{-k d}}{k d} \right)
       \sin^{2} \theta \cos^{2} \varphi \right],
\label{eq:S2}
\end{align}
where $A$ is the exchange stiffness, $M_{s}$ is the saturation magnetization, $k$ is the spin-wave wavenumber, $d$ is the film thickness, and $(\theta, \varphi)$ are the polar and azimuthal angles of the magnetization.

\begin{figure}[h] 
    \centering
    \includegraphics[width=\linewidth]{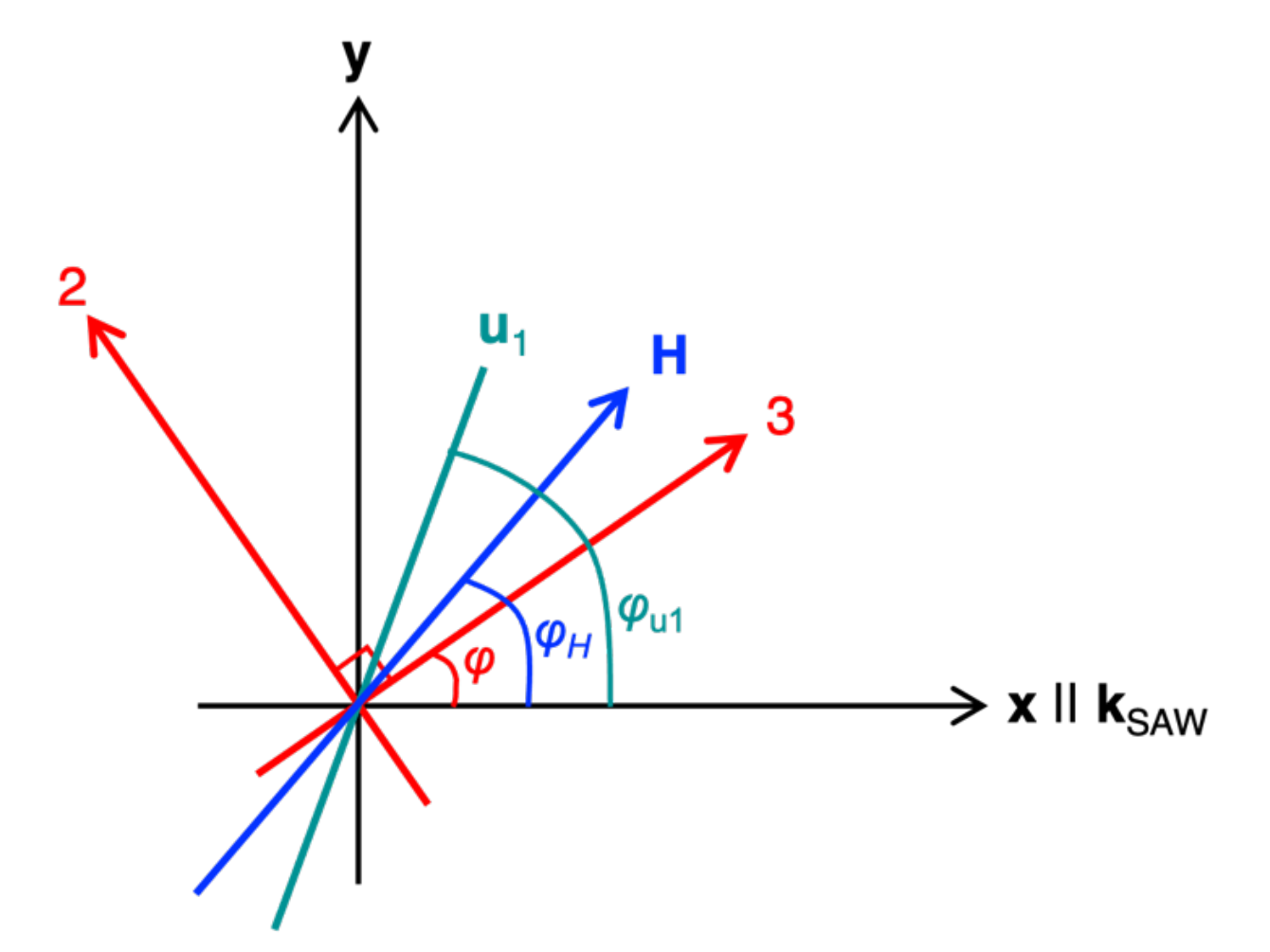}
    \caption{The $(1,2,3)$ coordinate system and the anisotropy unit vector 
    $\mathbf{u}_1$ in the $xy$ plane. The 1-axis is perpendicular to the page. 
    With respect to the $x$ axis, $\varphi$, $\varphi_H$, and $\varphi_{u_1}$ 
    are the angles of magnetization, $\mathbf{H}$, and $\mathbf{u}_1$, respectively. 
    For clarity, $\mathbf{u}_2$ is not shown in this coordinate system; 
    however, $\mathbf{u}_2$ is also in the $xy$ plane, and its angle from the $x$ axis 
    is $\varphi_{u_2}$.}
    \label{fig:coordinate_system}
\end{figure}

We employ the $(1,2,3)$ coordinate system in which the 3-axis is along $\mathbf{m}$ and the 2-axis lies in the film plane, as introduced by Dreher \textit{et al.}~\cite{Dreher2012}. The transformation between the $(x,y,z)$ and $(1,2,3)$ coordinates is
\begin{align}
\begin{pmatrix}
m_{x} \\[2pt]
m_{y} \\[2pt]
m_{z}
\end{pmatrix}
&=
U
\begin{pmatrix}
m_{1} \\[2pt]
m_{2} \\[2pt]
m_{3}
\end{pmatrix},
\label{eq:S3}
\end{align}
with
\begin{align}
U &=
\begin{pmatrix}
\cos\theta_{0} \cos\varphi_{0} & -\sin\varphi_{0} & \sin\theta_{0} \cos\varphi_{0} \\
\cos\theta_{0} \sin\varphi_{0} &  \cos\varphi_{0} & \sin\theta_{0} \sin\varphi_{0} \\
-\sin\theta_{0}                 & 0               & \cos\theta_{0}
\end{pmatrix},
\label{eq:S4}
\end{align}
where $\theta_{0}$ and $\varphi_{0}$ are the polar and azimuthal equilibrium angles of $\mathbf{m}$.

The vector components in the $(1,2,3)$ coordinate system are defined as
\begin{align}
\mathbf{H} &= (H_{1}, H_{2}, H_{3}), &
\mathbf{m} &= (m_{1}, m_{2}, m_{3}), \nonumber\\
\mathbf{u}_{1} &= (u_{1,1}, u_{1,2}, u_{1,3}), &
\mathbf{u}_{2} &= (u_{2,1}, u_{2,2}, u_{2,3}).
\label{eq:S5}
\end{align}

Using Eqs.~(\ref{eq:S3}) and (\ref{eq:S4}), the free energy [Eq.~(\ref{eq:S2})] can be rewritten as
\begin{align}
G &= - \mu_{0} \mathbf{H} \cdot \mathbf{m}
   + B_{d} (-m_{1} \sin\theta_{0} + m_{3} \cos\theta_{0})^{2} \nonumber\\
  &\quad - B_{u1} (\mathbf{m} \cdot \mathbf{u}_{1})^{2}
     - B_{u2} (\mathbf{m} \cdot \mathbf{u}_{2})^{4} \nonumber\\
  &\quad + \frac{A}{M_{s}} k^{2}
     + \frac{\mu_{0} M_{s}}{2} 
       \left[ \frac{1 - e^{-k d}}{k d} \cos^{2} \theta \right. \nonumber\\
  &\quad\quad
       \left. + \left( 1 - \frac{1 - e^{-k d}}{k d} \right)
         \sin^{2} \theta \cos^{2} \varphi \right],
\label{eq:S6}
\end{align}
where we have neglected the small $A$ term for simplicity.

Define
\begin{align}
\eta &= \frac{1 - e^{-k d}}{k d}.
\label{eq:S7}
\end{align}

Since $\mathbf{u}_{1}$ and $\mathbf{u}_{2}$ are in-plane, $u_{1,1} = u_{2,1} = 0$. From Fig.~\ref{fig:coordinate_system},
\begin{align}
u_{1,2} &= \sin(\varphi - \varphi_{u_{1}}), &
u_{1,3} &= \cos(\varphi - \varphi_{u_{1}}), \nonumber\\
u_{2,2} &= \sin(\varphi - \varphi_{u_{2}}), &
u_{2,3} &= \cos(\varphi - \varphi_{u_{2}}).
\end{align}

When $\theta_{0} = 90^{\circ}$, the second derivatives are
\begin{align}
G_{11} &= 2 B_{d}, \label{eq:S8}\\
G_{12} &= 0, \label{eq:S9}\\
G_{22} &= - 2 B_{u1} \sin^{2}(\varphi - \varphi_{u_{1}}) \nonumber\\
       &\quad - 12 B_{u2} \sin^{2}(\varphi - \varphi_{u_{2}})
              \cos^{2}(\varphi - \varphi_{u_{2}}) \nonumber\\
       &\quad + \mu_{0} M_{s} (1 - \eta) \sin^{2} \varphi, \label{eq:S12}\\
G_{3}  &= - \mu_{0} H \cos(\varphi - \varphi_{H}) \nonumber\\
       &\quad - 2 B_{u1} \cos^{2}(\varphi - \varphi_{u_{1}}) \nonumber\\
       &\quad - 4 B_{u2} \cos^{4}(\varphi - \varphi_{u_{2}}).
\label{eq:S13}
\end{align}

\section{PARAMETERS FOR CALCULATIONS}
The material and magnetic parameters adopted in the simulations are summarized in Table~\ref{tab:params}. The values are consistent with previously reported ranges~\cite{Hwang2024,Hatanaka2022,Gao2022}.

\begin{table}[h] 
    \caption{Parameters used for calculations.}
    \label{tab:params}
    \begin{ruledtabular}
    \begin{tabular}{llc}
    Symbol & Description & Value \\
\hline
$\gamma$ & Gyromagnetic ratio & $1.92 \times 10^{11}\,\mathrm{Hz/T}$ \\
$B_{1}, B_{2}$ & Magnetoelastic coupling constants & $14\,\mathrm{T}$ \\
$\epsilon_{xx}$ & Longitudinal strains & $1 \times 10^{-6}$ \\
$\epsilon_{xy}$ & Shear strains & $1 \times 10^{-7}$ \\
$\alpha$ & Gilbert damping factor & $0.1$ \\
$M_{s}$ & Saturation magnetization & $2.5 \times 10^{5}\,\mathrm{A/m}$ \\
$B_{d}$ & Out-of-plane shape anisotropy & $\mu_{0}M_{s}/2$ \\
$B_{ui}$ & In-plane biaxial anisotropy & $2K_{i}/\mu_{0}M_{s}$ \\
\hline
& & \\
Device & $B_{u1}$ (mT) & $B_{u2}$ (mT) \\
& $\varphi_{1}$ (deg) & $\varphi_{2}$ (deg) \\
\hline
Device 1 & 0.76 & 0.25 \\
         & 145  & 55 \\
Device 2 & 0.89 & 0.38 \\
         & 60   & -30 \\
    \end{tabular}
    \end{ruledtabular}
\end{table}

\newpage
\bibliography{refs.bib}

\end{document}